%% file: polarised_zz.tex
\documentclass[a4paper,10pt]{article}
\pdfoutput=1

\usepackage{jheppub}
\usepackage{amssymb}
\usepackage{graphicx}
\usepackage[dvipsnames,table]{xcolor}
\usepackage{slashed}
\usepackage{hyperref}
\usepackage{xspace}
\usepackage[subrefformat=parens, position=top, skip=-15pt, margin=15pt, justification=justified, singlelinecheck=false] {subcaption}
\usepackage{amsmath}
\usepackage{amsfonts}
\usepackage{mathrsfs}
\usepackage{comment}
\usepackage{afterpage}
\usepackage{verbatim}
\usepackage{booktabs}
\usepackage{array}
\usepackage[title,titletoc]{appendix}
\usepackage{tabularx}
\usepackage{colortbl, xcolor}
\usepackage{soul}
\usepackage[normalem]{ulem}
\newcolumntype{C}[1]{>{\centering\arraybackslash}p{#1}}

\newcommand{\nnb}{\nonumber}
\newcommand{\mc}{\mathcal}

\input{macros.tex}

\newcommand{\process}{\Pp\Pp\to\Pe^{+}\Pe^{-}\mu^{+}\mu^{-}\Pj\Pj}

\title{The cleanest of them all: NLO electroweak corrections to vector-boson scattering into doubly polarised $\PZ\PZ$ pairs at the LHC}
\author[a]{Ansgar Denner,}
\author[a]{Robert Franken,}
\author[a]{Santiago Lopez Portillo Chavez,}
\author[b]{Daniele Lombardi,}
\author[c]{Giovanni Pelliccioli}
\affiliation[a]{Institut f\"ur Theoretische Physik und Astrophysik,
  Universit\"at W\"urzburg, \\Emil-Hilb-Weg~22, 97074 W\"urzburg, Germany}
\affiliation[b]{Dipartimento di Fisica, Universit\`a  di Torino and INFN, Sezione di Torino, \\
Via P. Giuria 1, 10125 Torino, Italy}
\affiliation[c]{Universit\`a degli Studi di Milano--Bicocca
  and INFN Sezione di Milano--Bicocca, \\
  Piazza della Scienza 3, 20126 Milano, Italy}
\emailAdd{ansgar.denner@uni-wuerzburg.de}
\emailAdd{robert.franken@uni-wuerzburg.de}
\emailAdd{santiago.lopez-portillo-chavez@uni-wuerzburg.de}
\emailAdd{daniele.lombardi@unito.it}
\emailAdd{giovanni.pelliccioli@unimib.it}

\abstract{We present the first calculation of the
  next-to-leading-order electroweak
  corrections to vector-boson scattering into doubly polarised
  Z~bosons at the LHC in the fully leptonic decay channel.
  The production and decay of the two polarised Z~bosons are
  consistently modelled in the
  double-pole approximation, separating polarisation states at the
  amplitude level and including factorisable real and virtual
  electroweak corrections. Doubly polarised and unpolarised signals
  are investigated and confronted with off-shell results. 
  A broad analysis, including results at integrated and differential level,
  is carried out in a realistic, CMS-inspired fiducial
  setup. Our study paves the way to upcoming analyses with LHC Run-3
  and High-Luminosity data as well as to 
further phenomenological investigations.}%

\preprint{COMETA-2026-21}

\begin{document}
\strut\hfill\draftdate\\
\maketitle
\flushbottom

\section{Introduction}
\label{se:intro}
Vector-boson scattering (VBS) plays a key role in probing the electroweak (EW) and scalar sectors of the Standard Model (SM) and, in particular, the mechanism of electroweak-symmetry breaking.
Its sensitivity to quartic gauge couplings, which is enhanced by strong gauge cancellations,
renders it a powerful laboratory to test the consistency of the SM in the high-energy regime as well as an important probe of beyond-the-SM effects.

Among the various VBS channels accessible at the Large Hadron Collider (LHC),
the production of two $\PZ$~bosons in association with two jets occupies a special position.
When the $\PZ$~bosons decay to charged leptons, the four-lepton final state leads to a remarkably clean experimental signature.
At variance with $\PW\PZ$ and $\PW\PW$ production mechanisms, which involve one or two neutrinos in the final state,
in $\PZ\PZ$ production all final-state particles can be reconstructed, enabling a precise determination of the event kinematics.
Thus, $\PZ\PZ$ scattering with four final-state charged leptons is the golden channel for detailed studies of VBS dynamics and especially
the analysis of polarisation-sensitive observables.
At the same time, the four-lepton channel suffers from a rather small
cross section, owing to the small branching ratio of two $\PZ$~bosons into charged leptons.
As a consequence, the experimental measurement is statistically dominated, but an improvement is expected with the upcoming High-Luminosity LHC run, hopefully allowing
for polarisation analyses.
From the theoretical viewpoint, precise and accurate predictions for $\PZ\PZ$ scattering require
the inclusion of higher-order perturbative corrections in the EW and the strong (QCD) coupling,
a proper treatment of off-shell and interference effects, and the application of realistic fiducial selections
to enable a meaningful comparison between real LHC data and numerical SM predictions.

Experimentally, the $\PZ\PZ$ scattering into four charged leptons has been investigated \cite{CMS:2017zmo,CMS:2020fqz} and observed \cite{ATLAS:2020nlt,ATLAS:2023dkz}
with LHC Run-2 data. The low statistics makes the analysis of the polarisation structure of this channel intricate, but prospect studies for the upcoming High-Luminosity upgrade are promising \cite{CMS:2018mbt}.

On the theory side, the next-to-leading-order (NLO) QCD corrections in the SM to both the signal and irreducible background have been known for a long time
\cite{Jager:2006cp,Campanario:2014ioa} and also matched to a
parton shower \cite{Jager:2013iza}. The modelling of the gluon-initiated loop-induced irreducible background has been studied
as well \cite{Li:2020nmi}. The EW corrections as well as the full tower of NLO corrections to off-shell $\PZ\PZ$ scattering at the LHC 
have been calculated for both the signal and the irreducible-background processes \cite{Denner:2020zit,Denner:2021hsa}. All known predictions pertain the final-state signature with four leptons and two jets, 
\ie the fully leptonic decay channel, which is also considered in this work.
The SM predictions for polarised $\PZ$~bosons in $\PZ\PZ$ scattering are currently limited to
leading order (LO) combined with parton-shower simulations via matching and merging techniques
in the \phantommc \cite{Ballestrero:2019qoy}, \madgraphnlo \cite{BuarqueFranzosi:2019boy},
and \Sherpa \cite{Hoppe:2023uux} Monte Carlo frameworks.

In this work, we analyse the polarisation structure of $\PZ\PZ$ scattering in the decay channel with two different-flavour charged-lepton pairs,
achieving for the first time NLO accuracy in the EW coupling to both production and decay of polarised $\PZ$~bosons.

This paper is organised as follows. In \refse{se:calculation} we
describe the considered process, show the results of the validation of our calculation, and detail input parameters and event selections.
The integrated and differential results are presented in \refse{se:results} for a realistic fiducial setup. We draw our conclusions in \refse{se:conclusion}.

\section{Details of the calculation}
\label{se:calculation}

We consider the LHC signature
\begin{equation}\label{eq:full_process}
\process,
\end{equation}
which includes VBS into Z bosons decaying leptonically.
The off-shell modelling of this process has been
studied in \citeres{Denner:2020zit,Denner:2021hsa}, where a detailed presentation of
the contributing partonic channels and of sample Feynman diagrams was given.
We restrict ourselves here to the LO
and NLO EW terms, \ie the $\order{\alpha^6}$ and
$\order{\alpha^7}$ perturbative orders. The dominant contributions to the cross section originate from
quark--quark-induced channels involving quarks of the first two
generations. We do not take into account bottom-quark channels, 
whose impact is at the level of a few percent \cite{Denner:2020zit}.
Similarly, photon--photon-induced channels are disregarded.
At LO, these $\gamma\gamma$-induced channels contribute at the level of parts-per-million.
We do, however, include single-photon-induced channels ($\gamma q$), which involve
contributions with vector-boson-scattering topologies at $\order{\alpha^7}$.

While we compute the full off-shell process in
\refeq{eq:full_process}, the main target of this work are the
\emph{on-shell} signals with either polarised or unpolarised intermediate
\PZ~bosons, \ie
\begin{equation}\label{eq:DPA_process}
\Pp\Pp\to\PZ_{\lambda_1}(\to\Pe^+\Pe^-)\,\PZ_{\lambda_2}(\to\mu^+\mu^-)\Pj\Pj,
\qquad \lambda_1,\lambda_2 = \rL,\rT,\rU,
\end{equation}
where the indices $\rL$, $\rT$, and $\rU$ label longitudinal,
transverse, and unpolarised Z~bosons. We stress that the transverse
polarisation includes the left--right interference terms, as
usually done in experimental analyses that rely on polarised-template fits.

The calculation of the full off-shell process uses the complex-mass scheme
\cite{Denner:1999gp,Denner:2005fg,Denner:2006ic,Denner:2019vbn} to
treat the vector-boson resonances in a gauge-invariant way. On the other hand, gauge invariance for the on-shell signal processes in
\refeq{eq:DPA_process} is guaranteed by making use of the
pole approximation \cite{Stuart:1991xk,Aeppli:1993rs,Denner:2000bj,Denner:2019vbn}.
Within this  approximation, the polarised signals are defined as
specified in \citeres{Denner:2020bcz,Denner:2021csi,Denner:2024tlu},
where all relevant analytical results can be found.
The polarisation states of the massive vector bosons are frame
dependent. We choose to define them in the di-boson centre-of-mass
frame, which is best motivated from a theoretical point of view
\cite{Denner:2024tlu} and used in recent experimental analyses \cite{%
CMS:2020etf, 
ATLAS:2025wuw}.
Non-factorisable EW corrections, which are expected to be suppressed in
the phase-space region dominated by the Z-boson resonances, are neglected.

The subtraction of the infrared (IR) divergences is based on the
Catani--Seymour dipole formalism
\cite{Catani:1996vz,Nagy:1998bb,Dittmaier:1999mb,Catani:2002hc,Dittmaier:2008md,Basso:2015gca}.
IR singularities originating from the splitting of virtual
photons into quark--antiquark pairs are regularised using the
photon-to-jet conversion function \cite{Denner:2019zfp}.

The calculation within the double-pole approximation (DPA) requires a tailored dipole selection,
which we carry out using the general strategy 
outlined in \citeres{Denner:2024tlu,Denner:2026phn} restricted to the case of neutral resonances.

The final state of \refeq{eq:DPA_process} receives contributions
involving three vector-boson resonances, which become sizeable if the invariant mass of the
jet pair is close to a vector-boson mass. These contributions are
singular in the DPA, where all resonance widths are set to zero in the 
residue of the amplitude. While at LO such contributions are
removed by the cut on the jet-pair invariant mass,
\refeq{eq:invmasscut}, this cut can be evaded if a
real-emission photon is recombined with one of the jets in radiative
events. For the regularisation of these kind of singularities, we use
the \emph{fudge-factor scheme} introduced in \citere{Denner:2025xdz}
and available in \mocanlo \cite{Denner:2026phn,denner_2026_19829093} and \bbmc. To this
end, for each
potential resonance that is not treated within the pole approximation,
the squared matrix element is multiplied by a fudge factor, defined as
\begin{align}\label{eq:fudge_factor}
\mathcal{F}(s_V)=
\begin{cases}
  1 & |\sqrt{s_V}-M_V|>c_{\mc F}\,\Gamma_V\\
  \frac{(s_V-M^2_V)^2}{(s_V-M^2_V)^2+\Gamma^2_VM^2_V} &  |\sqrt{s_V}-M_V|<c_{\mc F}\,\Gamma_V\\
  \end{cases},
\end{align}
where  $s_V$ refers to the Lorentz invariant associated with the unprotected resonance $V$
($=\PW,\,\PZ$) decaying to a quark--antiquark pair, and $c_{\mc F}$ is
some parameter, for which we have chosen $c_{\mc F}=3$.
Since the fudge factors multiply complete gauge-invariant squared matrix elements,
gauge invariance is not violated.

\subsection{Tools and Validation}

The calculation has been performed with the Monte Carlo integration
codes \mocanlo\cite{Denner:2026phn,denner_2026_19829093} and \bbmc. While both use \recola
\cite{Actis:2012qn,Actis:2016mpe} and \collier
\cite{Denner:2016kdg} for the evaluation of LO and NLO matrix
elements, they rely on independent implementations of the
phase-space integration based on multi-channel importance sampling
according to
\citeres{Berends:1994pv,Denner:1999gp,Dittmaier:2002ap}.

As a validation of our calculation, all numerical results for cross sections and distributions reported in \refse{se:results}
have been produced with both integration codes, showing agreement at the few permille level.
The results for the different contributions of the employed subtraction formalism,
i.e.~integrated dipole, real and virtual contributions, have been compared and separately
agree within the statistical uncertainty of the calculation.

The results presented in \refse{se:results} are based on \mocanlo.

\subsection{Input parameters}
\label{se:setup}

We have generated results for
the Run-3 energy of the LHC, \ie $\sqrt{s} =
13.6\TeV$. The on-shell values for masses and widths of EW bosons
are taken from the PDG review \cite{ParticleDataGroup:2024cfk},
\begin{align}\label{eq:EWmasses}
  \MZOS    &{}=  91.1880\GeV,\quad & 
  \GZOS    &{}=  2.4955  \GeV,\nonumber \\ 
  \MWOS    &{}=  80.3692 \GeV,\quad & 
  \GWOS    &{}=  2.085  \GeV, 
\end{align}
and are converted into pole values ($\MZ,\MW,\GW,\GZ$) following
\citere{Bardin:1988xt}, leading to
\begin{align}
  \MZ &=  91.153872568\GeV,       & \quad \quad \quad \GZ &= 2.494566050\GeV, \\
  \MW &=  80.342168302\GeV,       & \GW &= 2.084298723\GeV.
\end{align}

The parameters for the top quark and the Higgs boson read \cite{ParticleDataGroup:2024cfk}
\begin{align}
\Mt &{}= 172.57\GeV,\quad & 
\Gt &{}= 1.42\GeV, \nnb\\  
\MH &{}= 125.20\GeV,\quad & 
\GH &{}= 0.0041\GeV,  
\end{align}
where the Higgs width is taken from \citere{LHCHiggsCrossSectionWorkingGroup:2016ypw}.
Leptons and light quarks, including the bottom quark, are treated as
massless.

The EW coupling is fixed via the $G_\mu$ input scheme \cite{Denner:2000bj,Dittmaier:2001ay}.
For the off-shell calculation this amounts to using
\begin{equation}\label{eq:alphaOFFSH}
  \alpha =
  \frac{\sqrt{2}}{\pi}\,G_\mu
  \,\left|\mu_{\PW}^2\,
  \left(  1 - \frac{\mu_{\PW}^2}{\mu_{\PZ}^2} \right)
  \right|\,,
  \qquad \mu_V^2=M_V^2-\ri M_V\Gamma_V\,, \quad V=\PW,\,\PZ\,,
\end{equation}
with the complex masses $\mu_V$ in the complex-mass scheme
\cite{Denner:2005fg,Denner:2006ic}.
Instead, the (un)polarised results in DPA are obtained with the
coupling computed from real masses
\begin{equation}\label{eq:alphaONSH}
  \alpha =
  \frac{\sqrt{2}}{\pi}\,G_\mu\,\MW^2\,
  \left(
  1 - \frac{\MW^2}{\MZ^2}
  \right).
\end{equation}
In both cases, the Fermi constant is set to
\begin{align}
\GF = 1.1663788 \times 10^{-5}\GeV^{-2}\,.
\end{align}
Thus, we use the values resulting from \refeq{eq:alphaOFFSH} and \refeq{eq:alphaONSH},
\begin{equation}
  \alpha = 1/132.159951034\,,\qquad
  \alpha = 1/132.223957653\,,
\end{equation}
for the off-shell and DPA calculations respectively.

We employ the \texttt{NNPDF40\_nnlo\_as\_01180\_qed}~\cite{NNPDF:2024djq} PDF set
and dynamical renormalisation and factorisation scales,
\begin{equation}
\label{eq:scale}
\mu_{\rm R} =
\mu_{\rm F} =
\sqrt{\,\pt{\Pj_1}\,\pt{\Pj_2}} \,,
\end{equation}
where $p_{\rT\Pj_1}$ and $p_{\rT\Pj_2}$ are the transverse momenta of
the two jets with highest transverse momenta.

\subsection{Event selection}\label{sec:setup}
The event selection is adapted from the CMS note \cite{CMS:2018mbt}, in combination with the lepton selection cuts from \citere{CMS:2016ogx}.
All partons with pseudorapidity $|\eta| > 5$ are assumed to be lost in the beam pipe. For the remaining partons, we apply a two-step recombination procedure.
First, we dress leptons with photons with a resolution parameter $R = 0.1$ in the Cambridge--Aachen algorithm \cite{Dokshitzer:1997in}.
Then, quarks, gluons, and the remaining photons are clustered into jets with a resolution parameter $R = 0.4$ using the anti-$k_{\rm T}$ algorithm \cite{Cacciari:2008gp}.

We require both $\PZ$~bosons to decay into lepton pairs
of different generations, which results in exactly one electron, one
positron, one muon, and one antimuon.  The four leptons are ordered
according to their transverse momenta and labelled $\Pl_1$ to $\Pl_4$.
Their respective transverse momenta have to exceed
\begin{align}
\pt{\Pl_1} > 20\GeV, \qquad \pt{\Pl_2} > 10\GeV, \qquad \pt{\Pl_3} >
5\GeV, \qquad \pt{\Pl_4} > 5\GeV.
\end{align}
The pseudorapidities of all four leptons have to fulfil
\begin{align}
|\eta_{\Pl}| < 2.5.
\end{align}
Each same-family lepton--antilepton pair must be inside an invariant mass range of
\begin{align}\label{eq:invmasscut}
60\GeV < M_{\Pl^+\Pl^-} < 120\GeV,
\end{align}
and the invariant mass of the four-lepton system is bounded by
\begin{align}
M_{4\Pl} > 180\GeV.
\end{align}
Only jets with transverse momentum and pseudorapidity
\begin{align}
\pt{\Pj} > 30\GeV, \qquad |\eta_\Pj| < 4.7
\end{align}
are considered. The two hardest jets, also called tagging jets, need to fulfil
\begin{align}
|\Delta \eta_{\Pj_1\Pj_2}| > 2.4, \qquad  M_{\Pj_1\Pj_2}>400 \GeV.
\end{align}
Furthermore, we require a separation between the tagging jets and each lepton of
\begin{align}
\Delta R_{\Pj_1\Pl_k}, \Delta R_{\Pj_2\Pl_k} > 0.4, \quad \forall k\in\{1,\dots, 4\},
\end{align}
where $\Delta R_{ij} = \sqrt{(\Delta \phi_{ij})^2 + (\Delta \eta_{ij})^2}$ (with $\Delta \phi$ as the azimuthal-angle difference).

\section{LHC phenomenology}
\label{se:results}
\subsection{Integrated results}
In this section we present the integrated cross sections for the polarised and unpolarised $\PZ\PZ$-scattering process defined in \refeqs{eq:full_process}--\eqref{eq:DPA_process} in the fiducial setup described in \refse{sec:setup}.
In \refta{tab:NLO_xsec} we show fiducial cross sections at LO
[$\order{\alpha^6}$] and NLO EW order [$\order{\alpha^7}$],
highlighting the relative impact of quark-induced and photon-induced contributions to the various polarisation states, as well as to the unpolarised process.
\begin{table}
\centering
\begin{tabular}{c|cccccc}
        mode &         $\sigma_\mathrm{LO}^{\alpha^6}$ $[\ab]$ &
                       $\Delta \sigma_{\mathrm{NLO},\Pq}^{\alpha^7}$ $[\ab]$ & $\delta_{\mathrm{NLO},\Pq}$ [\%] &
                       $\Delta \sigma_{\mathrm{NLO},\gamma}^{\alpha^7}$ $[\ab]$ & $\delta_{\mathrm{NLO}, \gamma}$ [\%] &                         
                       $\sigma_\mathrm{NLO}^{\alpha^6 + \alpha^7}$ [ab] \\ \hline
%
         full         & $119.83(1)$  & $-21.4(1)$   & $-17.9$  & $1.850(5)$   & $1.5$  & $100.3(1)$  \\ 
         unp.         & $117.32(2)$  & $-20.52(1)$  & $-17.5$  & $1.7985(2)$  & $1.5$  & $98.59(2)$  \\ 
         LL           & $8.9677(2)$  & $-1.288(3)$  & $-14.4$  & $0.12329(3)$ & $1.4$  & $7.802(4)$  \\ 
         LT           & $16.6406(2)$ & $-2.715(4)$  & $-16.3$  & $0.24794(4)$ & $1.5$  & $14.173(9)$ \\ 
         TT           & $72.8492(4)$ & $-13.48(1)$  & $-18.5$  & $1.1488(1)$  & $1.6$  & $60.51(1)$  \\ 
         int.         & $2.22(2)$    & $-0.32(2)$   & $-$      & $0.0305(2)$  & $-$    & $1.93(3)$   \\ 
\end{tabular}
\caption{Fiducial integrated cross sections (in $\ab$ units) at LO
  [$\order{\alpha^6}$] (2nd column) and NLO EW corrections
  [$\order{\alpha^7}$] for $\PZ\PZ$ scattering at the LHC. The NLO EW
  corrections are listed in absolute terms and relative to the LO in the
  third and fourth columns for the quark-induced contributions only,
  and in the fifth and sixth columns for the single-photon-induced
  contributions. Owing to symmetric cuts on the leptons, the TL
  contribution is identical to the LT one and thus not shown. The seventh column
  reports the sum of the two computed orders.
}\label{tab:NLO_xsec}
\end{table}
The relative NLO EW corrections are defined via
\begin{equation}
\delta_{\mathrm{NLO},i} = \frac{\Delta \sigma_{\mathrm{NLO},i}^{\alpha^7}}{\sigma_\mathrm{LO}^{\alpha^6}}.
\end{equation}
While the single-photon-induced contributions are at the $1.5\%$ level and
rather independent of the vector-boson polarisation,
the corrections to quark-induced partonic processes are
driven by the leading EW Sudakov logarithms
in the one-loop virtual corrections. The EW Casimir factors multiplying the leading and sub-leading logarithms
depend on the polarisation mode of the EW bosons \cite{Denner:2000jv}, therefore leading to different EW corrections with a similar hierarchy
as in other VBS production mechanisms \cite{Denner:2024tlu,Denner:2025xdz}. In particular,
the more longitudinal vector bosons are involved, the smaller the
negative EW corrections are.

As for the polarisation fractions, defined as
\beq\label{eq:polfrac}
f_{k,\lambda\lambda'} = \frac{\sigma_{k,\lambda\lambda'}}{\sigma_{k,\rm unp.}}\,,\qquad \lambda,\lambda'=\rL,\rT,\qquad k ={\rm LO,\,NLO}\,,
\eeq
results similar to those in $\PW^+\PW^+$
\cite{Denner:2024tlu} and $\PW^+\PZ$ scattering \cite{Denner:2025xdz} are found,
with NLO EW corrections only mildly modifying the corresponding LO predictions.
The numerical results are shown in
Table~\ref{tab:NLO_polfrac}.
\begin{table}
\centering
\begin{tabular}{c|cc}
  mode &  $f_\mathrm{LO}$ [\%]  & $f_\mathrm{NLO}$ [\%] \\
  \hline
         full          & $102.1$    & $101.7$ \\ 
         unp.          & $100.0$    & $100.0$ \\ 
         LL            & $7.6$      & $7.9$   \\ 
         LT            & $14.2$     & $14.4$  \\ 
         TT            & $62.1$     & $61.4$  \\ 
         int.          & $1.9$      & $1.9$      
\end{tabular}
\caption{Fiducial polarisation fractions at LO and NLO EW accuracy, defined as ratios of (un)polarised integrated cross sections over the unpolarised DPA one (unp.).}\label{tab:NLO_polfrac}
\end{table}
The purely longitudinal component amounts to about $8\%$ of the unpolarised cross section, 
the mixed ones (summed) to $28\%$, while the transverse one gives the
dominant contribution with more than $60\%$.
This hierarchy is in line with the expected high-energy behaviour in the SM. There, the longitudinal-boson 
scattering is subject to strong unitarity cancellations between energy-growing pure-gauge and Higgs-mediated contributions. 
The transverse modes benefit from a larger number of unsuppressed helicity configurations.
On the basis of previous results \cite{Denner:2024tlu}, there are good reasons to believe
that QCD corrections do not change dramatically the picture of
polarisation fractions. Of course, this could only be 
verified by computing these corrections, which goes beyond the scope of this work.

By subtracting the sum of polarised cross sections from the DPA
unpolarised one, we obtain an interference contribution, 
\beq\label{eq:int}
\sigma_{k,\rm int.} = \sigma_{k,\rm unp.}-
\sum_{\lambda,\lambda'=\rL,\rT}\sigma_{k,\lambda\lambda'}\,,
\qquad k ={\rm LO,\,NLO}\,,
\eeq
which roughly amounts to a positive $2\%$ both at LO and NLO EW.
Finally, the difference between the off-shell calculation and the DPA unpolarised one gives genuine 
off-shell effects (beyond the pole approximation) at the 
$2\%$ level both at LO and NLO EW.

\subsection{Differential distributions}
Before showing differential results, we remark that the analysis of $\PZ\PZ$ scattering at the LHC in the decay channel 
with four charged leptons is clearly statistically dominated.  
In the considered fiducial setup \cite{CMS:2018mbt}, about $300$ (unpolarised) signal events are expected with the full High-Luminosity
dataset ($3000\fb^{-1}$), translating into about $20$ $\rL\rL$ events. This makes the extraction of the $\rL\rL$ fraction challenging already 
at the level of the fiducial measurement. In spite of the signal purity, differential measurements will only be possible by combining different decay channels.
Nonetheless, the differential description of $\PZ\PZ$ scattering in the four-charged-lepton channel is useful to further 
understand the process under the lenses of polarisations and spin correlations. 
Moreover, the purity of this decay channel makes it possible to directly access a plethora of kinematic observables.

In the considered setup, the kinematic selections applied to the charged leptons are symmetric, therefore the two mixed polarisation states 
give equivalent contributions. 
More precisely, the $\rT\rL$ contribution to an observable depending
on some electron and muon properties is equal to the $\rL\rT$ contribution 
to the observable with electron and muon properties interchanged. 
Therefore, in all figures we only show distributions for the $\rL\rT$
state.    

\begin{figure}[hbt]
\centering
\begin{subfigure}{0.49\textwidth}
\centering
\includegraphics[page=1,width=1.\linewidth]{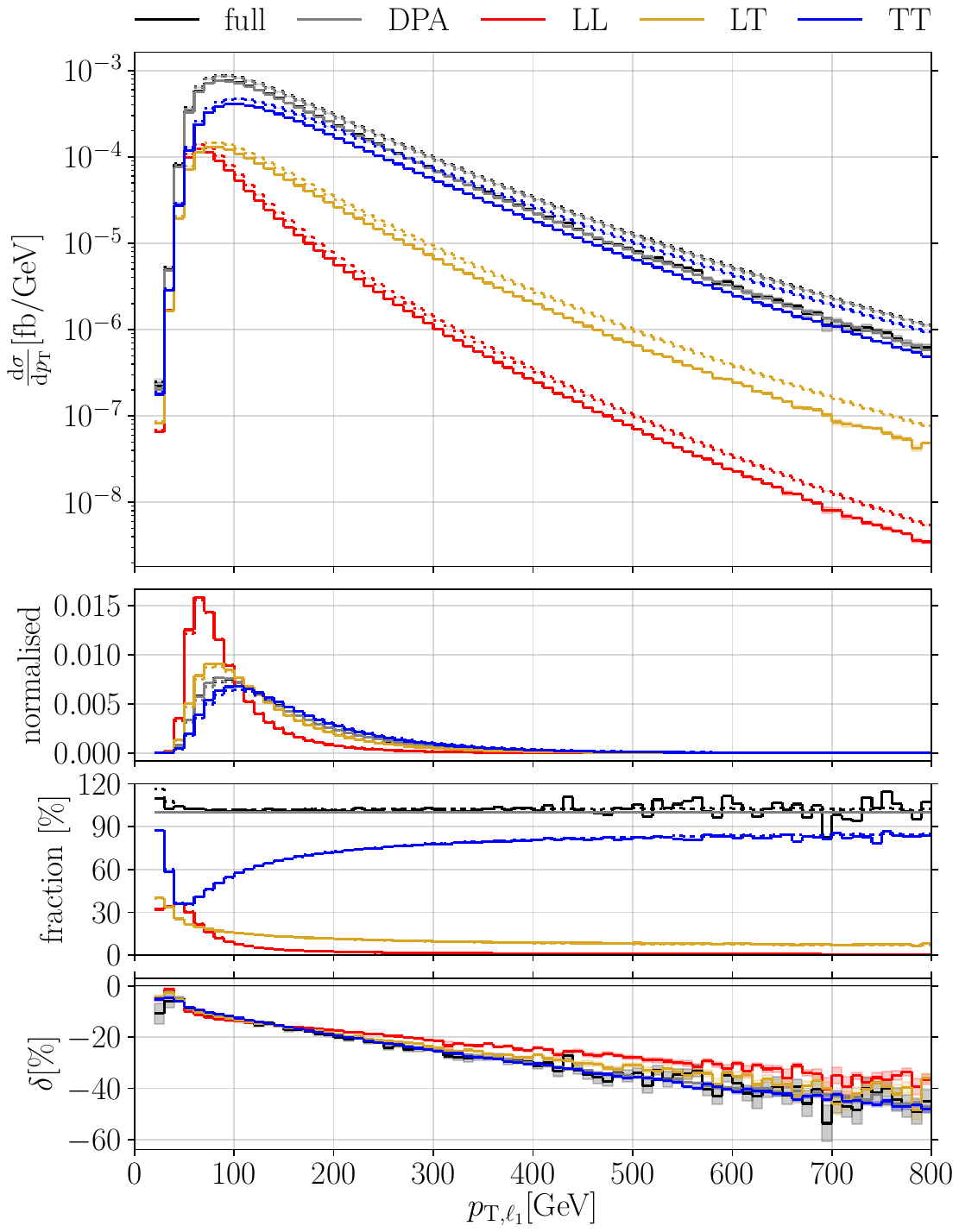}
\end{subfigure}
\begin{subfigure}{0.49\textwidth}
\centering
\includegraphics[page=2,width=1.\linewidth]{plots/mocanlo_NLO_plot_log.pdf}
\end{subfigure}
\caption{Differential cross section with respect to the transverse
  momentum of the hardest lepton (left) and the (vectorial) sum of
  transverse momenta of the four leptons (right). The figure is
  organised as follows. The top panel shows absolute cross sections, while
  the second one normalises the cross sections such that the area
  under all curves equals one. The third panel displays the fraction
  of the polarisation contribution with respect to the unpolarised DPA
  cross section. The bottom panel depicts the relative NLO EW
  corrections in percentages ($\delta$). Solid lines denote the NLO cross section, dashed lines
  the LO one. Shaded bands indicate the numerical-integration error.}
\label{fig:distr_pt_e_pt_ZZ}
\end{figure}

We start by presenting transverse-momentum distributions of the
leading charged lepton and of the system of four charged leptons in \reffi{fig:distr_pt_e_pt_ZZ}.
The two observables share many similar features.
In the moderate- and large-$p_{\rT}$ regime of both observables, all doubly polarised states decrease at the same rate, 
with a clear hierarchy in the absolute size lead by the $\rT\rT$ state,
which is approximately one order of magnitude larger than the mixed states and two orders of magnitude larger than the $\rL\rL$ state.
In the tails, all LO distributions get increasing negative EW corrections,
dominated by large EW Sudakov logarithms appearing in the one-loop SM amplitudes.
These negative corrections, which reach up to $-40\%$ at $\pt{\PZ\PZ}=800\GeV$, are slightly larger 
for the transverse modes of the two $\PZ$~bosons, owing to a larger EW Casimir factor multiplying 
the leading Sudakov logarithms compared to the purely longitudinal state.
In this region, however, the cross section is roughly three orders of magnitude smaller than in the 
most populated region. Both transverse-momentum distributions show a marked difference between 
the purely longitudinal state and all other states in the normalised shapes. 
Indeed, leptons from longitudinally polarised bosons are preferably emitted
orthogonal to the direction of the boson, whereas for transversely polarised bosons the
decay products tend to be aligned or anti-aligned with it. Owing to this fact, the $\rL\rL$ state 
populates more the soft part of the $p_{\rm T}$ spectra, while being more suppressed in the 
moderate-$p_{\rm T}$ region. This effect, already found in other VBS production mechanisms 
\cite{Denner:2024tlu,Denner:2025xdz}, makes these transverse-momentum observables suitable 
for polarisation discrimination, even with a reduced number of bins.

We continue with angular observables associated with the decay leptons, which are expected to inherit in a more direct manner the spin structure of the
underlying resonances (the $\PZ$~bosons) compared to energy-dependent quantities.
\begin{figure}
\centering
\begin{subfigure}{0.49\textwidth}
\centering
\includegraphics[page=12,width=1.\linewidth]{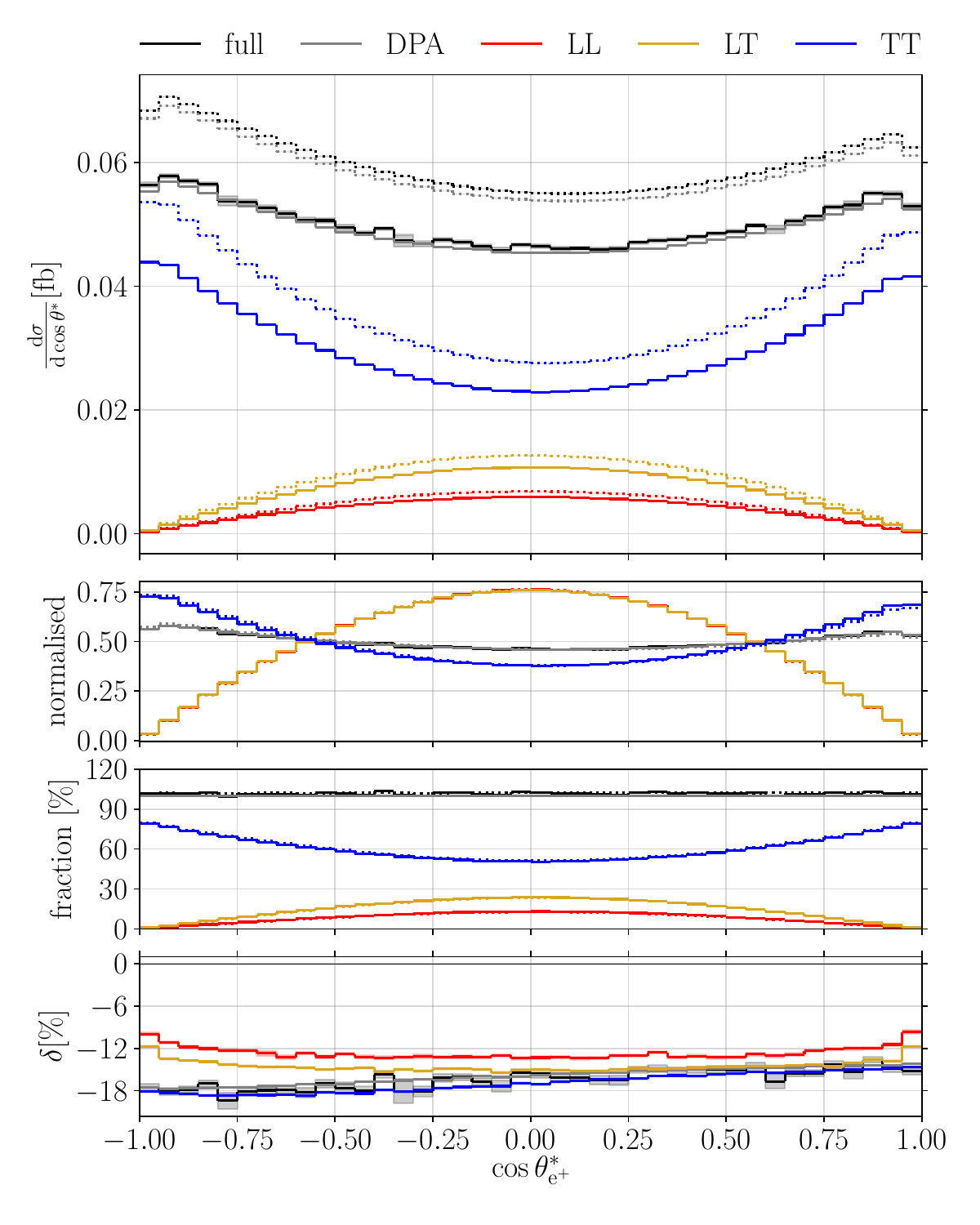}
\end{subfigure}
\begin{subfigure}{0.49\textwidth}
\centering
\includegraphics[page=27,width=1.\linewidth]{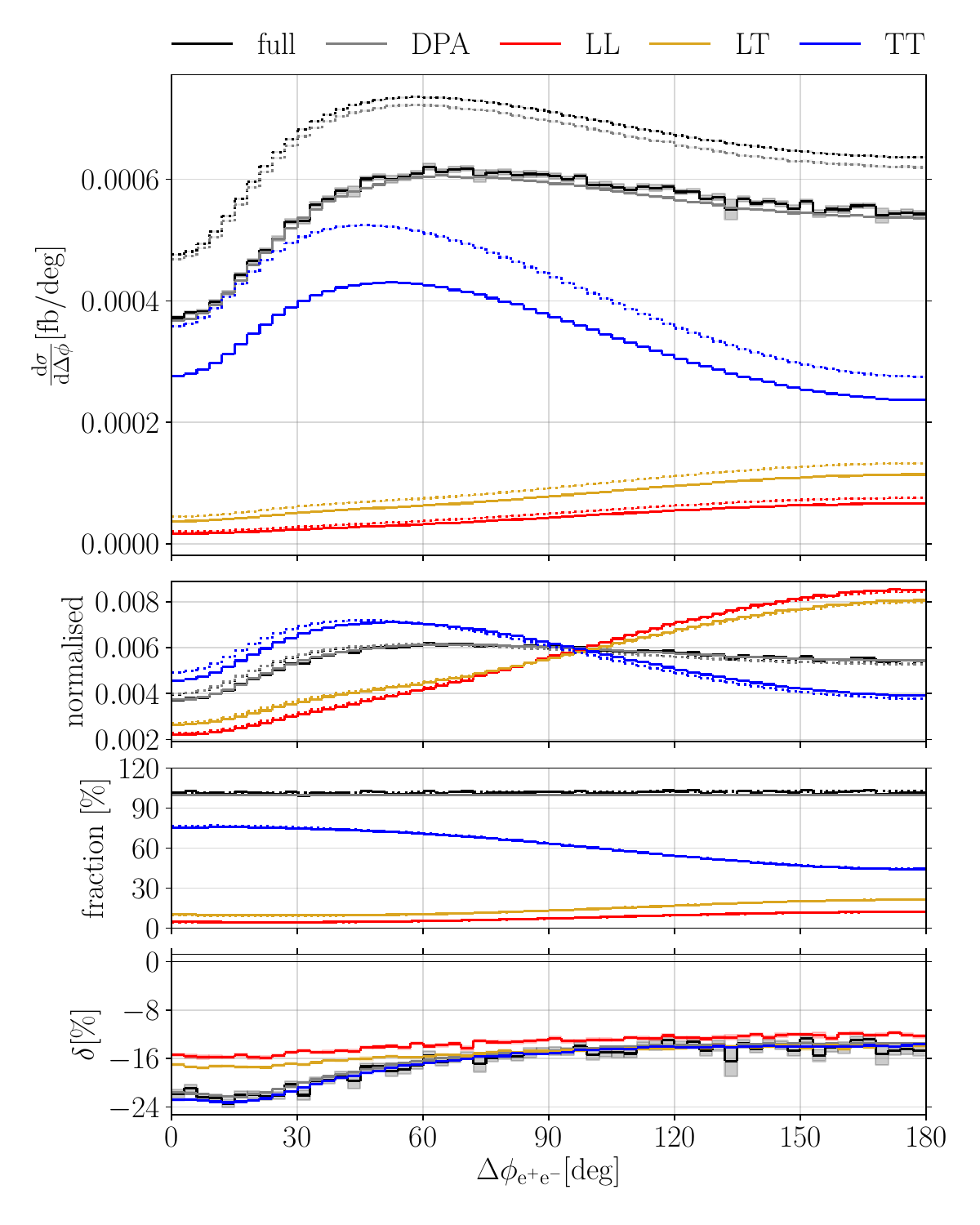}
\end{subfigure}
\caption{Differential cross section with respect to the cosine of the positron decay angle in the corresponding $\PZ$-boson rest frame (left)
and to the azimuthal-angle distance between the electron and the positron (right). Same structure as \reffi{fig:distr_pt_e_pt_ZZ}.}
\label{fig:distr_cost_ep_Dphi_ee}
\end{figure}
In \reffi{fig:distr_cost_ep_Dphi_ee} we consider the distribution in
the cosine of the polar decay angle of the positron in the corresponding $\PZ$-boson rest frame,
which is the quantity that is most sensitive to the polarisation state of the individual $\PZ$~boson (the one decaying to the positron--electron pair)
out of the considered observables.
At variance with the analogous angle in $\PW\PZ$ scattering (see Fig.~5(a) in \citere{Denner:2025xdz}), the reconstruction of the positron decay angle only relies on
the kinematics of visible particles. Therefore, up to small distortions at the endpoints owing to the loose cuts,
the (normalised) distributions show the expected quadratic behaviour with opposite-sign second derivative (negative for $\rL\rL$ and $\rL\rT$, positive
for $\rT\rL$ and $\rT\rT$) \cite{Bern:2011ie,Stirling:2012zt}. The EW corrections are almost flat and reflect the results at the integrated level.

The distribution in the azimuthal-angle distance (computed in the laboratory frame) between the positron and the electron, presented in the right panels of 
\reffi{fig:distr_cost_ep_Dphi_ee}, is strongly correlated with the positron decay angle and therefore possesses a marked discrimination power
for the polarisation of the single $\PZ$~boson. The EW corrections to this quantity are rather flat for $\Delta\phi_{\Pe^+\Pe^-}>90^\circ$. 
Below this value they reach up to $-25\%$ for the transverse modes ($\rT\rT$, but also $\rT\rL$ which is not shown) for $\Delta\phi_{\Pe^+\Pe^-}\to 0$,
while they remain at the $-15\%$ level for the longitudinal modes. This behaviour results from the correlation of high-energy $\PZ$~bosons with small $\Delta\phi_{\Pe^+\Pe^-}$.
In fact, smaller $\Delta\phi_{\Pe^+\Pe^-}$ values result from stronger boosted
Z~bosons, whose higher energy leads to larger EW Sudakov corrections.

\begin{figure}
\centering
\begin{subfigure}{0.49\textwidth}
\centering
\includegraphics[page=42,width=1.\linewidth]{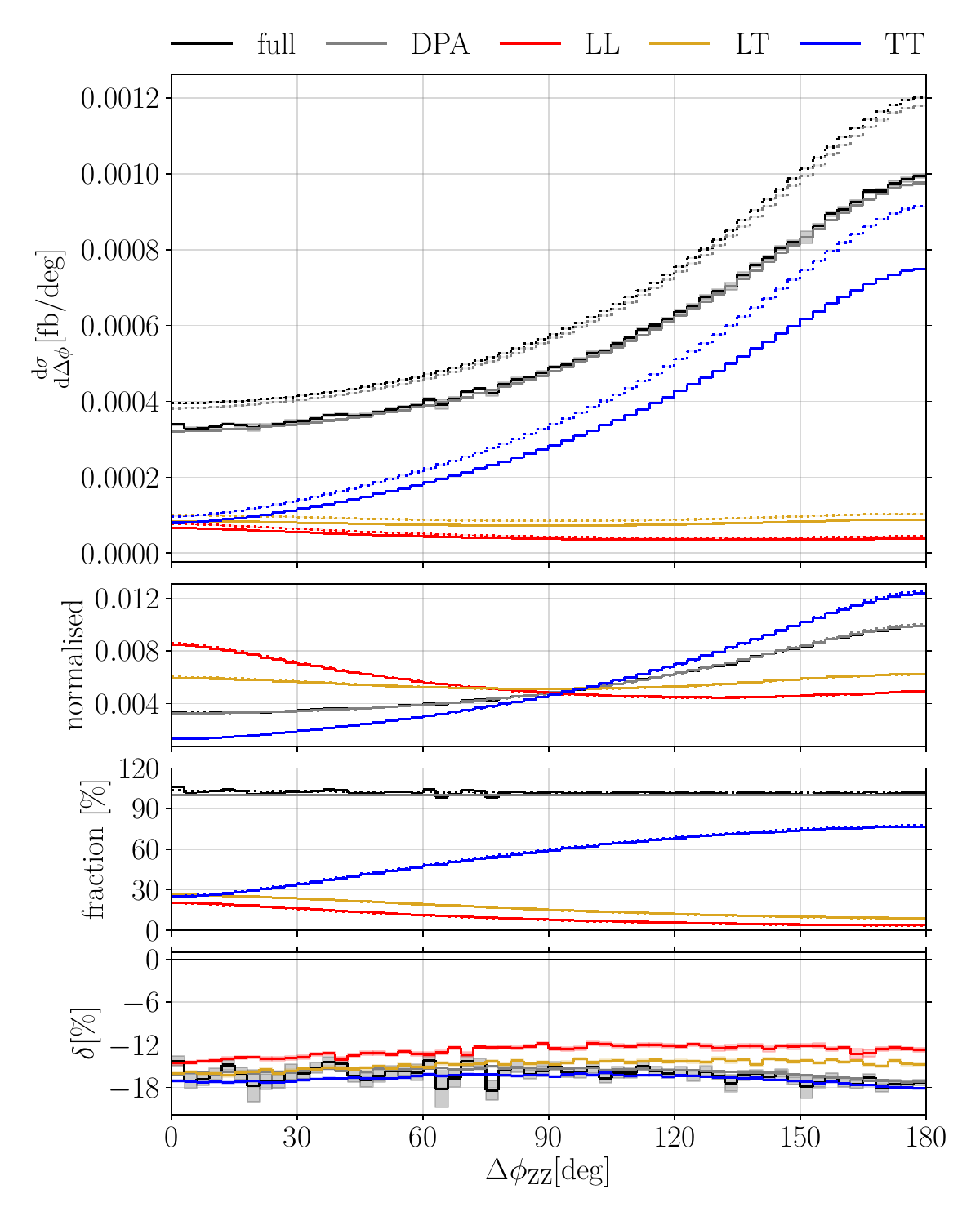}
\end{subfigure}
\begin{subfigure}{0.49\textwidth}
\centering
\includegraphics[page=52,width=1.\linewidth]{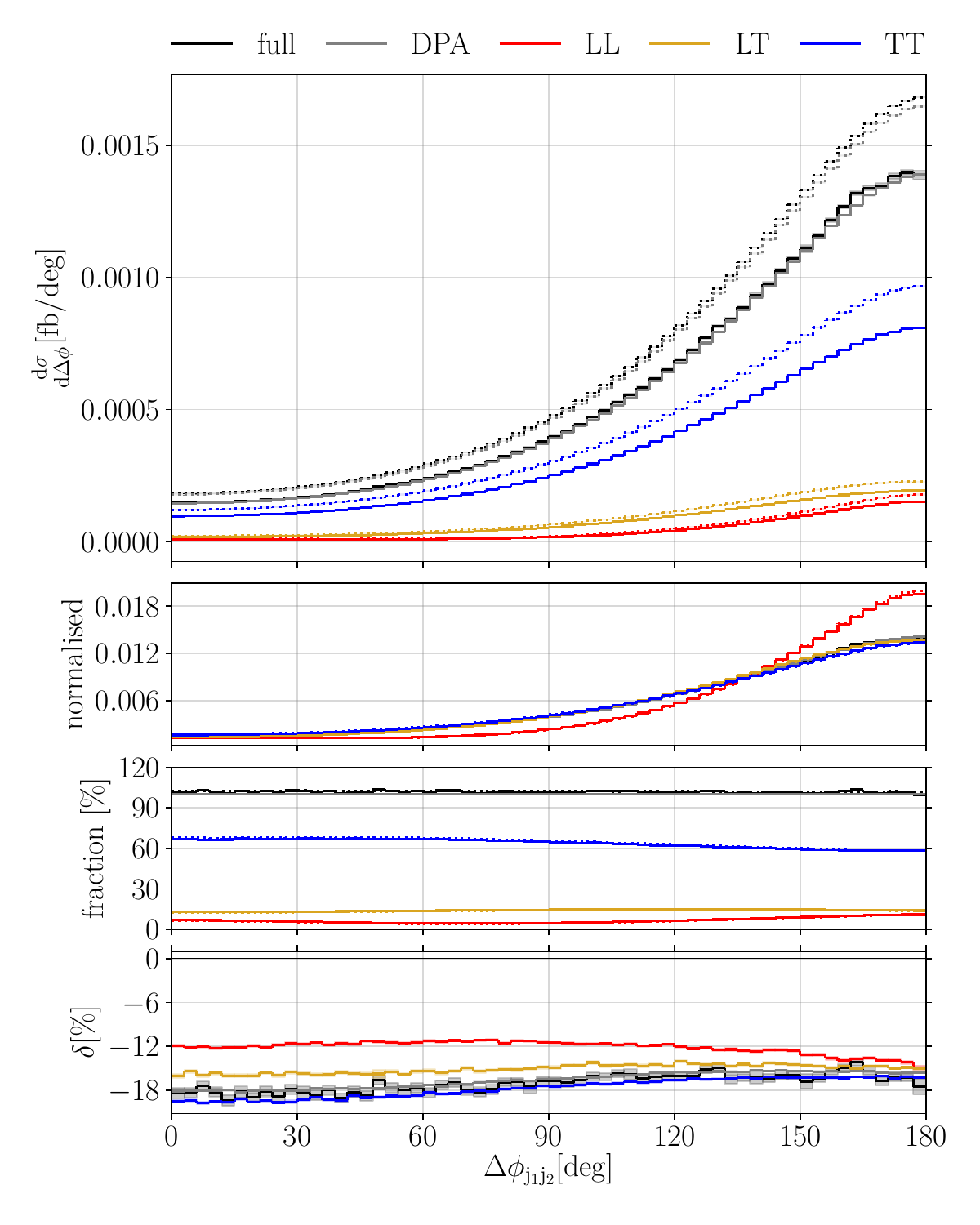}
\end{subfigure}
\caption{Differential cross section with respect to
  the azimuthal-angle distance between the two reconstructed $\PZ$~bosons (left)
  and between the two tagging jets (right). Same structure as
  \reffi{fig:distr_pt_e_pt_ZZ}.
}\label{fig:distr_Dphi_ZZ_jj}
\end{figure}

In \reffi{fig:distr_Dphi_ZZ_jj} we consider two observables defined as the azimuthal-angle distance between two objects.
The one on the left is associated with the two $\PZ$~bosons (identified as the two same-flavour lepton pairs).
While two longitudinal bosons are preferably produced in the same hemisphere, two transverse bosons are dominantly produced in
opposite hemispheres, leading to a marked normalised-shape difference between the two states. The mixed polarisation states
show a rather flat shape, symmetric about $\Delta\phi_{\PZ\PZ}=90^\circ$.
The picture for the azimuthal-angle separation between the two tagging jets is different as shown on the right plot of \reffi{fig:distr_Dphi_ZZ_jj}. In this case, the two jets
are preferably produced in opposite hemispheres, as usual in processes dominated by vector-boson-fusion (VBF) topologies.
While this holds for all polarisation states, there is a remaining shape difference between the $\rL\rL$ state and
all others. This is a clear example of the marked correlation between the spin structure of the di-boson system
and the di-jet-system kinematics in VBF topologies.
The EW corrections are rather flat for both azimuthal-angle
separations. Nevertheless, the differences in EW corrections between the different
polarisation modes increase for large  $\Delta\phi_{\PZ\PZ}$ and small $\Delta\phi_{\Pj_1\Pj_2}$.

\begin{figure}
\centering
\begin{subfigure}{0.49\textwidth}
\centering
\includegraphics[page=25,width=1.\linewidth]{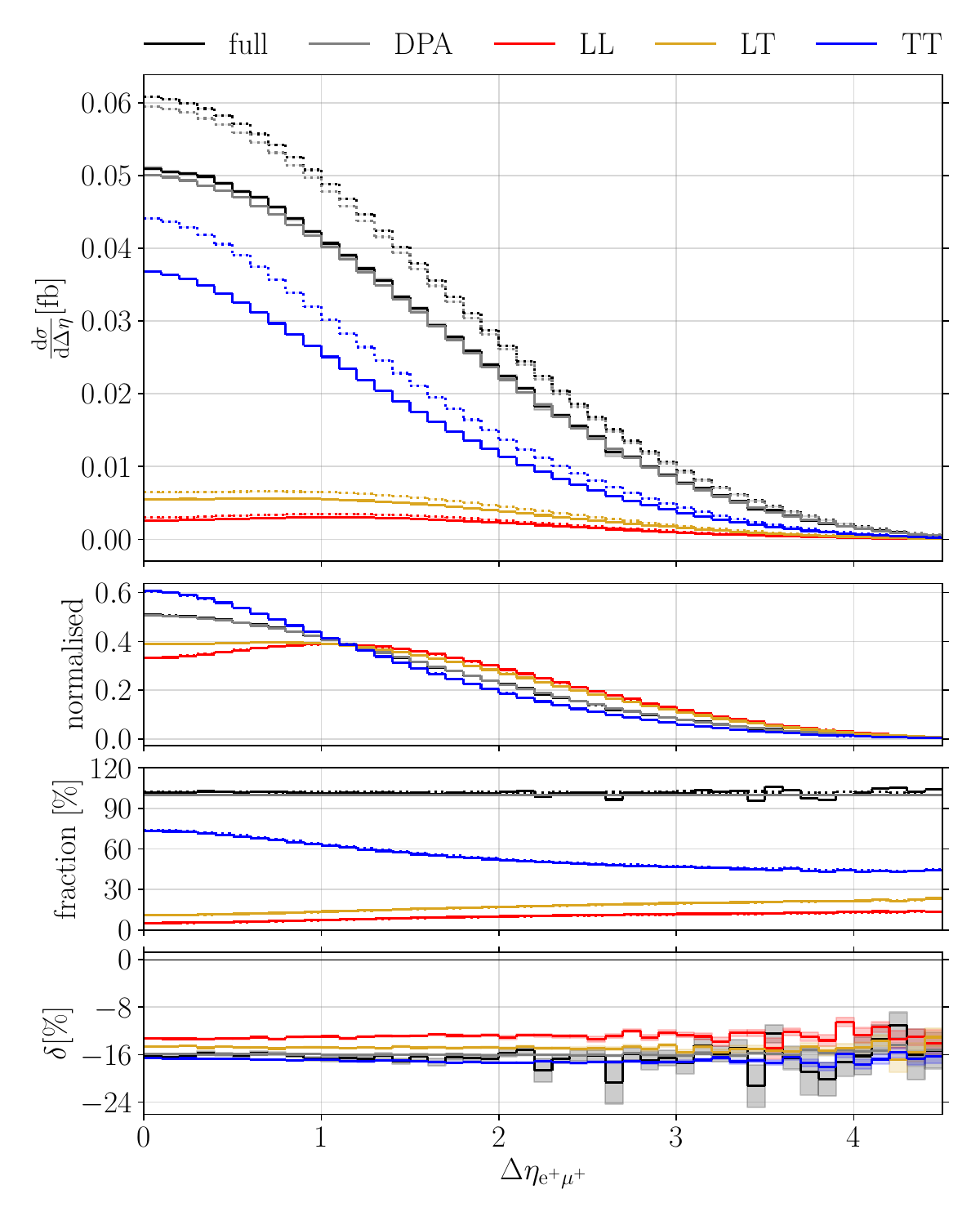}
\end{subfigure}
\begin{subfigure}{0.49\textwidth}
\centering
\includegraphics[page=44,width=1.\linewidth]{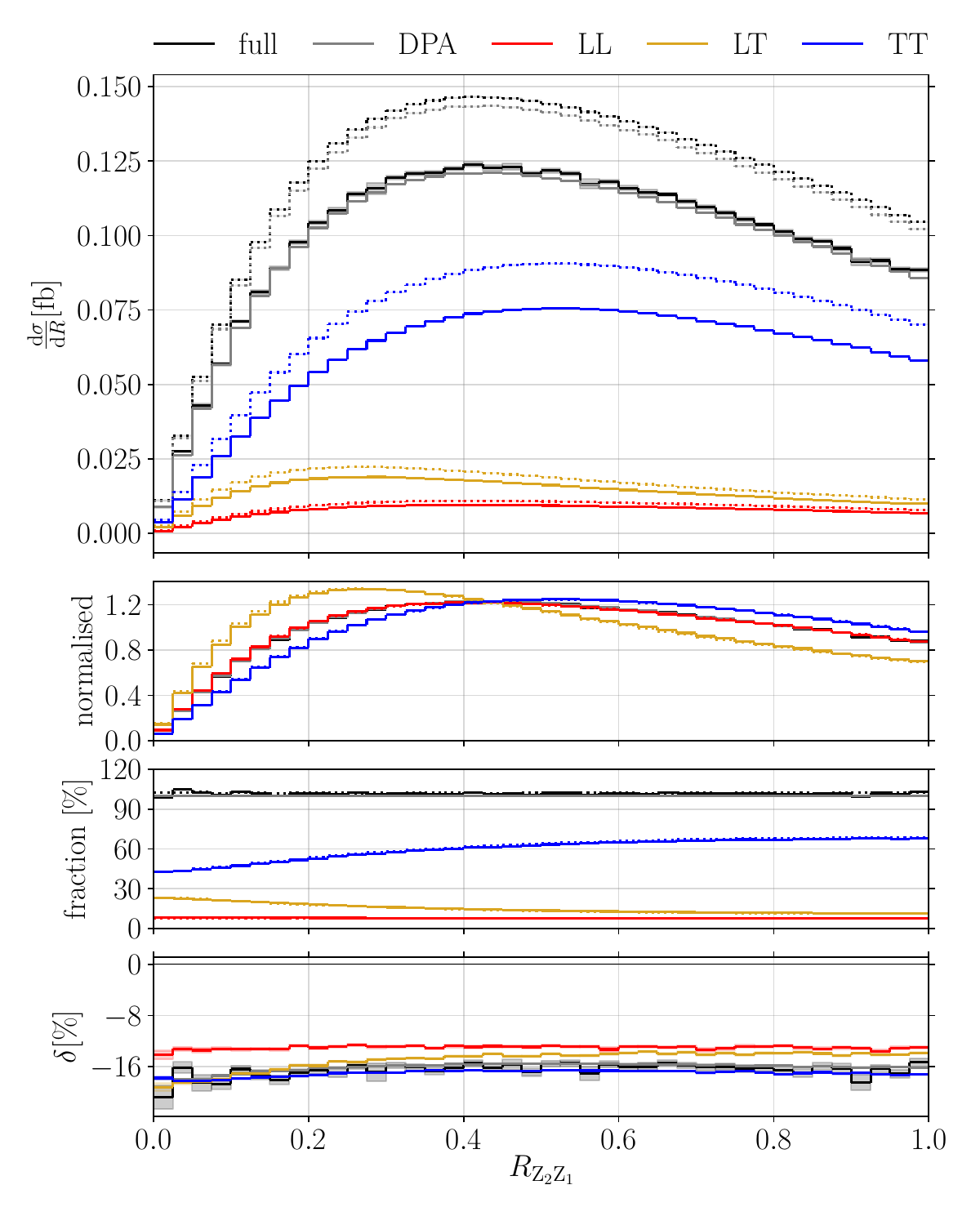}
\end{subfigure}
\caption{Differential cross section with respect to
  the rapidity difference between the positron and the antimuon (left) and to
  the transverse momentum ratio between the subleading and the leading reconstructed $\PZ$~boson (right).
  Same structure as \reffi{fig:distr_pt_e_pt_ZZ}. 
}
\label{fig:distr_R21_dy}
\end{figure}

In the left panels of \reffi{fig:distr_R21_dy} we show distributions in the
rapidity difference between the two positively charged leptons.
This quantity, already studied in the case of same-sign $\PW\PW$ production \cite{Denner:2024tlu},
has a fair discrimination power between doubly polarised signals. The $\rT\rT$ shape has a maximum
at zero rapidity separation, while the $\rL\rL$ one is maximal slightly above $\Delta\eta_{\Pe^+\mu^+}=1$.
The EW corrections do not sizeably change the LO distribution shapes.

As a last differential observable, we present in the right panels of \reffi{fig:distr_R21_dy}
the ratio between the transverse momenta of the sub-leading and
leading $\PZ$~bosons $R_{\PZ_2\PZ_1}=p_{\rT,\PZ_2}/p_{\rT,\PZ_1}$.
As also seen in $\PW\PZ$ scattering \cite{Denner:2025xdz}, the softer transverse-momentum spectrum associated with
longitudinal bosons compared to transverse ones leads to a ratio $R_{\PZ_2\PZ_1}$ that is typically smaller for the
$\rL\rT$ compared to the $\rT\rT$ state. 
The EW corrections to this quantity are rather flat for all polarisation states. 
Nevertheless, for modes involving transverse bosons a 
slope in the relative corrections is visible.

\section{Conclusion}
\label{se:conclusion}
We have achieved for the first time NLO EW accuracy
in the electroweak production of two polarised $\PZ$~bosons in association with two
jets at the LHC with subsequent decays to charged leptons.

Specific helicity states of intermediate $\PZ$~bosons
are defined at the level of SM tree-level and one-loop resonant amplitudes, working in the double-pole approximation to ensure both gauge invariance and a sound definition of polarised intermediate states.
The factorisable EW corrections  to the production and to the decay mechanisms from both real and virtual contributions
are consistently combined. This approach makes it possible to estimate both interference and genuine non-resonant effects 
by comparison with the unpolarised pole-approximated and the full off-shell calculations, respectively.

Large and polarisation-dependent EW corrections are found, ranging
from $-14\%$ ($\rL\rL$) to $-18\%$ ($\rT\rT$) in a realistic
CMS-inspired fiducial volume. A number of transverse-momentum and
angular observables show a marked
discrimination power between doubly polarised $\PZ\PZ$ signals and
only mild distortions of distribution shapes owing to EW corrections.

This work represents an important advancement for the interpretation of LHC Run-3 data
in the clean but statistically limited VBS signature with four charged leptons.
We remark that the numerical results shown in this paper can be reproduced with the upcoming 
public release of version 1.0.1 of the \mocanlo software,
\begin{center}
  \url{https://mocanlo.gitlab.io/releases/mocanlo-1.0.1.tar.gz}
\end{center}
using the input cards in the process subfolder
\begin{center}
 \texttt{mocanlo/validated\_processes/vbs/zz/pp\_zz\_ew\_ew\_pol/}
\end{center}

\section*{Acknowledgements}

The authors thank Sandro Uccirati and Jean-Nicolas Lang for the maintenance of
\recola and Claude Charlot, Roberto Covarelli, Lucia Di Ciaccio, Pietro Govoni, Joany Manjarres, Emmanuel Sauvan, Jean-Baptiste Sauvan for fruitful discussions.
AD and RF are supported by the German Federal Ministry for Education
and Research (BMBF) under contract no.~05H24WWA.
The research of DL has been supported by
the Italian Ministry of Universities and Research (MUR) under the FIS grant (CUP: D53C24005480001, FLAME).
GP acknowledges support from the EU Horizon Europe research and innovation programme
under the Marie-Sk\l{}odowska Curie Action
``POEBLITA - POlarised Electroweak Bosons at the LHC with Improved Theoretical Accuracy'' - grant agreement no.~101149251 (CUP H45E2300129000).

\bibliographystyle{JHEP}
\bibliography{pol}

\clearpage

\end{document}

%% file: macros.tex
\def\refeq#1{\mbox{Eq.~\eqref{#1}}}
\def\refeqs#1{\mbox{Eqs.~\eqref{#1}}}
\def\reffi#1{\mbox{Figure~\ref{#1}}}

\def\refta#1{\mbox{Table~\ref{#1}}}

\def\refse#1{\mbox{Section~\ref{#1}}}

\def\citere#1{\mbox{Ref.~\cite{#1}}}
\def\citeres#1{\mbox{Refs.~\cite{#1}}}

\newcommand{\newc}{\newcommand}
\newc{\beq}{\begin{equation}}
\newc{\eeq}{\end{equation}}
\newc{\beqn}{\begin{eqnarray}}
\newc{\eeqn}{\end{eqnarray}}
\newc{\bit}{\begin{itemize}}
\newc{\eit}{\end{itemize}}
\newc{\ben}{\begin{enumerate}}
\newc{\een}{\end{enumerate}}
\newc{\bce}{\begin{center}}
\newc{\ece}{\end{center}}
\newc{\bfi}{\begin{figure}}
\newc{\efi}{\end{figure}}

\newcommand{\ri}{\mathrm i}

\newcommand{\rT}{{\mathrm{T}}}

\newcommand{\rL}{{\mathrm{L}}}

\newcommand{\rU}{{\mathrm{U}}}

\newcommand{\ie}{{i.e.}\ }

\newcommand{\GeV}{\ensuremath{\,\text{GeV}}\xspace}
\newcommand{\TeV}{\ensuremath{\,\text{TeV}}\xspace}
\newcommand{\fb}{{\ensuremath\unskip\,\text{fb}}\xspace}
\newcommand{\ab}{{\ensuremath\unskip\,\text{ab}}\xspace}

\newcommand{\PH}{\ensuremath{\text{H}}\xspace}
\newcommand{\Pj}{\ensuremath{\text{j}}\xspace}
\newcommand{\Pp}{\ensuremath{\text{p}}}
\newcommand{\Pe}{\ensuremath{\text{e}}\xspace}
\newcommand{\Pm}{\ensuremath{\mu}\xspace}

\newcommand{\Pq}{\ensuremath{q}}
\newcommand{\Pt}{\ensuremath{\text{t}}\xspace}

\newcommand{\PW}{\ensuremath{\text{W}}\xspace}
\newcommand{\PZ}{\ensuremath{\text{Z}}\xspace}
\newcommand{\Pl}{\ell}
                                    
\newcommand{\pt}[1]{p_{\rT,{#1}}}

\newcommand{\Mt}{\ensuremath{m_\Pt}\xspace}
\newcommand{\MH}{\ensuremath{M_\PH}\xspace}
\newcommand{\MWOS}{\ensuremath{M_\PW^\text{OS}}\xspace}
\newcommand{\MW}{\ensuremath{M_\PW}\xspace}
\newcommand{\MZOS}{\ensuremath{M_\PZ^\text{OS}}\xspace}
\newcommand{\MZ}{\ensuremath{M_\PZ}\xspace}

\newcommand{\Gt}{\ensuremath{\Gamma_\Pt}\xspace}
\newcommand{\GH}{\ensuremath{\Gamma_\PH}\xspace}
\newcommand{\GZ}{\ensuremath{\Gamma_\PZ}\xspace}
\newcommand{\GZOS}{\ensuremath{\Gamma_\PZ^\text{OS}}\xspace}
\newcommand{\GW}{\ensuremath{\Gamma_\PW}\xspace}
\newcommand{\GWOS}{\ensuremath{\Gamma_\PW^\text{OS}}\xspace}

\newcommand{\GF}{\ensuremath{G_\mu}}

\newcommand{\order}[1]{\ensuremath{\mathcal{O}{\left(#1\right)}}\xspace}

\newcommand{\recola}{{\sc Recola}\xspace}

\newcommand{\Sherpa}{{\sc Sherpa}\xspace}

\newcommand{\mocanlo}{{\sc MoCaNLO}\xspace}
\newcommand{\bbmc}{{\sc BBMC}\xspace}
\newcommand{\collier}{{\sc Collier}\xspace}

\newcommand{\madgraphnlo}{{\sc\small MadGraph5\_aMC@NLO}\xspace}

\newcommand{\phantommc}{{\sc\small PHANTOM}\xspace}

\newcolumntype{.}{D{.}{.}{-1}}
\newcolumntype{d}[1]{D{.}{.}{#1}}
\colorlet{tableoverheadcolor}{gray!37.5}
\colorlet{tableheadcolor}{gray!25}
\colorlet{tablerowcolor}{gray!12.5}

\def\draftdate{\relax}
\def\mda{\relax}
\def\mua{\relax}
\def\mla{\relax}
\def\draft{
\def\thtystars{******************************}
\def\sixtystars{\thtystars\thtystars}
\typeout{}
\typeout{\sixtystars**}
\typeout{* Draft mode!
         For final version remove \protect\draft\space in source file *}
\typeout{\sixtystars**}
\typeout{}
\def\draftdate{\today}
\def\mua{\marginpar[\boldmath\hfil$\uparrow$]%
                   {\boldmath$\uparrow$\hfil}\color{black}%
                    \typeout{marginpar: $\uparrow$}\ignorespaces}
\def\mda{\color{red}\marginpar[\boldmath\hfil$\downarrow$]%
                   {\boldmath$\downarrow$\hfil}%
                    \typeout{marginpar: $\downarrow$}\ignorespaces}
\def\mla{\marginpar[\boldmath\hfil$\rightarrow$]%
                   {\boldmath$\leftarrow $\hfil}%
                    \typeout{marginpar: $\leftrightarrow$}\ignorespaces}
\def\Mua{\marginpar[\boldmath\hfil$\Uparrow$]%
                   {\boldmath$\Uparrow$\hfil}\color{black}%
                    \typeout{marginpar: $\uparrow$}\ignorespaces}
\def\Mda{\color{red}\marginpar[\boldmath\hfil$\Downarrow$]%
                   {\boldmath$\Downarrow$\hfil}%
                    \typeout{marginpar: $\downarrow$}\ignorespaces}
\def\Mla{\marginpar[\boldmath\hfil\textcolor{red}{$\Rightarrow$}]%
                   {\boldmath\textcolor{red}{$\Leftarrow $}\hfil}%
                    \typeout{marginpar: $\leftrightarrow$}\ignorespaces}
\overfullrule 5pt
\oddsidemargin 15mm
\marginparwidth 29mm
}
